\documentclass[aps,prl,amsmath,two column,amssymb,floatfixng,showpacs,
superscriptaddress,footinbib]{revtex4}
\usepackage[dvips]{graphics}
\usepackage{bm}
\usepackage{float}
\usepackage{epsfig}
\usepackage{enumerate}
\usepackage{amsmath}
\usepackage{color}
\usepackage{graphicx}
\usepackage[colorlinks=true,linktoc=page,linkcolor=red,citecolor=blue]{hyperref}

\newcommand\bea{\begin{eqnarray}}
\newcommand\eea{\end{eqnarray}}
\newcommand\beq{\begin{equation}}  
\newcommand\eeq{\end{equation}}

\begin{document}

\title{\titlename}
\date{\today}
\title{Interpolating between Space-like and Time-like Entanglement via Holography}
	\author{Carlos Nunez}
	\email{c.nunez@swansea.ac.uk}
		\affiliation{Department of Physics, Swansea University,
Swansea SA2 8PP, United Kingdom}
\author{Dibakar Roychowdhury}
\email{dibakar.roychowdhury@ph.iitr.ac.in}
\affiliation{Department of Physics, Indian Institute of Technology Roorkee
Roorkee 247667, Uttarakhand,
India}

\date{\today}
\begin{abstract}
We study entanglement entropy for slab-like regions in quantum field theories, using their holographic duals. We focus on the transition between space-like and time-like separations. By considering boosted subsystems in conformal and confining holographic backgrounds, we identify two classes of extremal surfaces--real ones (Type I) and complex surfaces (Type II). These interpolate between the usual Ryu–Takayanagi prescription and its time-like generalizations. We derive explicit expressions for the entanglement entropy in both conformal and confining cases and discuss their behaviour across phase transitions,  and null limits. The interpolation between Type I and Type II surfaces reveals  an analytic continuation of the extremal surface across the light-cone. Our analysis also finds the existence of a Ryu-Takayanagi surface (Type I) even for time-like separations in the confining field theory case.
\end{abstract}

\maketitle

%
%
%
\section{Introduction and General Idea}

The AdS/CFT conjecture \cite{Maldacena:1997re} introduced the idea that space may emerge from the strongly coupled dynamics of a Quantum Field Theory (QFT). One route to understanding this emergence is through Entanglement Entropy (EE), which can be computed holographically via the proposal of Ryu and Takayanagi \cite{Ryu:2006bv, Ryu:2006ef, Hubeny:2007xt}. Further developments in \cite{Swingle:2009bg, VanRaamsdonk:2010pw} suggested that spatial directions might themselves emerge from entanglement in the QFT.

Beyond standard EE, alternative entanglement measures have been proposed, such as pseudo-entropy—a generalisation involving post-selection and depending on both initial and final states \cite{Nakata:2020luh, Mollabashi:2021xsd}. This concept has its correlate in the area of Tensor Networks. For sample papers see \cite{Muller-Hermes:2012irk, Hastings:2014qqa, Lerose:2020fhd, Bou-Comas:2024pxf}. The cross-fertilisation between Tensor Networks ideas and Holography is promising.

Building on these ideas, the concept of time-like entanglement entropy (tEE) was introduced in \cite{Doi:2022iyj, Doi:2023zaf}. This quantity can be understood as an analytic continuation of canonical EE (for example, as computed via the replica trick). Indeed, in two-dimensional conformal field theories, tEE can be calculated through the analytic continuation of correlation functions of twist operators.

This topic has recently attracted significant attention, particularly in the holographic context. Several works that have informed  and influenced the present study include \cite{Heller:2024whi, Xu:2024yvf, Guo:2025pru, Das:2023yyl, Grieninger:2023knz, Chu:2019uoh, Afrasiar:2024lsi, Afrasiar:2024ldn, Milekhin:2025ycm, Li:2022tsv, Roychowdhury:2025ukl, Nunez:2025gxq, Giataganas:2025div}.

Most holographic studies of EE have focused on {\it space-like} separations (with some attention to null separations in \cite{Bousso:2014uxa}). The extension to time-like separations (and hence to tEE) opens new directions, such as the computation of Liu–Mezei-type central charges, the study of entanglement phase transitions, dimensional dependence, applications to confining or anisotropic theories, and the role of dualities \cite{Nunez:2025gxq, Giataganas:2025div}.

The analytic continuation in the field-theoretic definition of tEE has a gravitational counterpart. However, identifying the corresponding minimal surfaces in Lorentzian (time-like) settings is subtler than in the space-like case. A recurring conclusion from recent studies is that the bulk extremal surfaces—or their turning points—must be complex to account for time-like separations \cite{Heller:2024whi, Guo:2025pru}.

In this work, we propose a method to interpolate between space-like and time-like entanglement entropies (for slab-like regions) by computing the EE for boosted subsystems. Our approach accommodates space-like, time-like, and null separations. The above mentioned interpolation is subtle as the EE diverges for null separations as we discuss. We avoid this divergence by going into the complex plane. This provides a new perspective on the transition between spatial and temporal entanglement. We find two qualitatively distinct classes of extremal surface embeddings. For space-like (Type I) embeddings, the turning point is real, much like in the standard Ryu–Takayanagi prescription. In contrast, for time-like separations (Type II), the turning point generically becomes complex, and the radial variable also acquires a complex value, implementing the mechanism proposed in \cite{Heller:2024whi, Guo:2025pru} in general $d$-dimensional conformal field theories.

\begin{figure}
    \centering
    \includegraphics[width=0.5\linewidth]{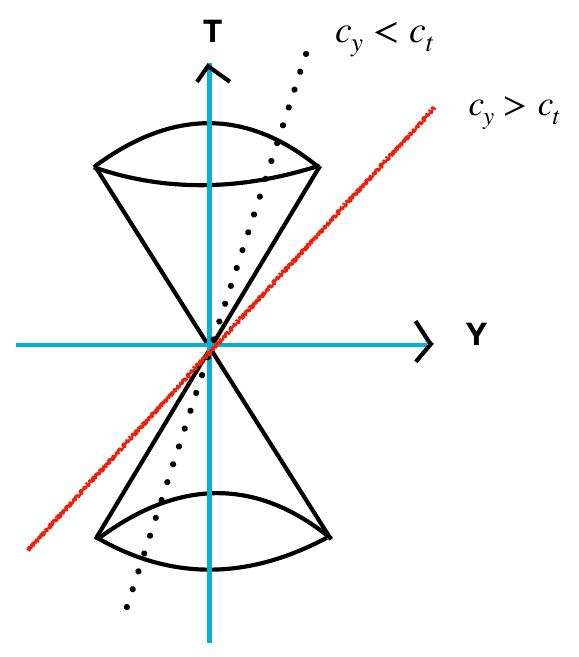}
    \caption{The red hyper-plane ($c_y>c_t$) corresponds to space-like separation in a boosted frame of reference. The black dotted hyper-plane ($c_y<c_t$) represents time-like separation in a boosted coordinate system.}
    \label{fig:light-cone}
\end{figure}

In fact, we compute EE for a slab region as a function of the invariant interval $\Delta^2 = Y^2 - T^2$. In Figure \ref{fig:light-cone} we see the two types of embeddings: Type I (for which the space separation is bigger than the time separation $Y>T$ and hence are space-like) and Type II (with $T>Y$).
We derive expressions for generic  conformal field theories in $d$-dimensional Minkowski space (which admit a holographic dual). These reduce to the standard Ryu–Takayanagi result \cite{Ryu:2006ef} for space-like separations, and to the analytically continued expressions of \cite{Doi:2023zaf} for time-like separations. Special cases—including AdS$_3$ and the null limit—are treated explicitly and emerge naturally from our formalism.

We also apply our method to a background interpolating between a four-dimensional CFT and a three-dimensional gapped and confining field theory \cite{Anabalon:2021tua}. We obtain the EE in terms of the slab interval, with both quantities expressed implicitly via the turning point of the extremal surface. Consistent with previous studies \cite{Nunez:2025gxq, Klebanov:2007ws, Kol:2014nqa}, we find a phase transition in the EE as a function of the slab interval—regardless of whether the separation is space-like or time-like. {In this  case, the confinement scale plays the role of a regulator, {\it allowing real-valued extremal surfaces even for time-like intervals}.  This is a nontrivial feature of the confining backgrounds we consider. It is similar to what occurs for black hole backgrounds, see Appendix in \cite{Bousso:2014uxa}.}

\underline{\bf Note added}: We learned that Michal Heller, Fabio Ori and Alexandre Serantes are working on similar ideas \cite{Heller:2025kvp}. We coordinated the submission to appear on the same day in arXiv.
 
\section{Calculation for a  generic CFT$_d$}

Let us consider a generic holographic dual to a CFT in $d$ space-time dimensions. In string frame, the ten dimensional  metric and the dilaton $\Phi$ read,
\begin{eqnarray}
   & & ds^2_{st}= f_1(\vec{v})AdS_{d+1} + g_{ij}(\vec{v})dv^i dv^j,~~\Phi=\Phi(\vec{v})\label{genericmetric}\\
   & & AdS_{d+1}= \frac{u^2}{l^2}(\lambda dt^2 + dy^2+ d\vec{x}_{d-2}^2)+\frac{l^2}{u^2} du^2.\nonumber
\end{eqnarray}
The metric $g_{ij}(\vec{v})$ refers to the internal space of dimension $(9-d)$. We introduced the parameter $\lambda=\pm 1$. When $\lambda=-1$ we are working in the Lorentzian metric, otherwise, we are working in an Euclidean situation. The background is complemented by Ramond and Neveu-Schwarz fields, that we do not quote.
To calculate the EE we consider an eight surface parameterised by the coordinates $\Sigma_8=[\vec{x}_{d-2}, u, \vec{v}_{9-d}]$, with $t(u), y(u)$. 
We calculate the induced metric on $\Sigma_8$ and the entanglement entropy. The results are,
\begin{eqnarray}
& & ds_{\Sigma_8,st}^2=  f_1(\vec{v})\left[ \frac{l^2}{u^2}du^2 \left( 1+\frac{u^4}{l^4} (y'^2 +\lambda t'^2)\right) +\frac{u^2}{l^2}d\vec{x}_{d-2}^2 \right] \nonumber\\
& & ~~+ ~~g_{ij, (9-d)}dv^i dv^j,\nonumber\\
& & 4G_{10} S_{EE}= {\cal N} \int_{u_0}^\infty \sqrt{G^2(u) + F^2(u) (y'^2+\lambda t'^2)}~ du,\label{EEgeneric}\\
& & \text{where}~~G= \left(\frac{u}{l}\right)^{d-3}, ~~~F=\left(\frac{u}{l}\right)^{d-1},~~~\text{and}\nonumber\\
& &{\cal N}= L_x^{d-2}\int d\vec{v} \sqrt{e^{-4\Phi(\vec{v})}\det g_{9-d}(\vec{v}) f_1(\vec{v})^{d-1}}.\nonumber
\end{eqnarray}
The quantity $\frac{{\cal N}}{4 G_{10}}$ is related to the central charge of the dual CFT. The important parameter $u_0$ is the turning point of the eight-surface, the lowest value of the coordinate $u$ reached by $\Sigma_8$. The equations of motion of the variables $t(u), y(u)$ are,
\begin{eqnarray}
& &    \frac{F^2 t'}{\sqrt{G^2+F^2 (\lambda t'^2 +y'^2)}}=\lambda c_t,\nonumber\\
& &
    \frac{F^2 y'}{\sqrt{G^2+F^2 (\lambda t'^2 +y'^2)}}= c_y.\nonumber
\end{eqnarray}
Here $c_y$ and $c_t$ are constants of motion.
The solution to these equations read
\begin{eqnarray}
 &   &t'^2(u)=\frac{c^2_t G^2(u)}{F^2(u)(F^2(u)-F^2(u_0))},\label{eq3}\\
  &  &y'^2 (u)=\frac{c^2_y G^2(u)}{F^2(u)(F^2(u)-F^2(u_0))},\nonumber
    \end{eqnarray}
where we denote the turning point
\begin{align}
    F^2(u_0)=c^2_y+\lambda c^2_t.
\end{align}
We can express  the lengths of the subsystems (in the coordinates $t$ and $y$ respectively)
\begin{eqnarray}
& &     T=2 c_t\int_{u_0}^{\infty}du \frac{G(u)}{F(u)\sqrt{F^2(u)-F^2(u_0)}},\nonumber\\
& &    Y=2 c_y\int_{u_0}^{\infty}du \frac{G(u)}{F(u)\sqrt{F^2(u)-F^2(u_0)}}.
\end{eqnarray}
The entanglement entropy after regularisation is,
\begin{align}
  &\frac{4G_{10}}{\cal N}  S_{EE}\nonumber\\
  &=2\int_{u_0}^{\infty}du \frac{G(u)F(u)}{\sqrt{F^2(u)-F^2(u_0)}}-2\int_{u_\ast}^\infty G(u)du.\label{regulatedEE}
\end{align}
The turning point is given by $ F^2(u_0)=c^2_y - c^2_t$ (we focus on the Lorentzian $\lambda=-1$ case in the following). The point $u_*$ denotes the end of the space, being $u_*=0$ for AdS-Poincare coordinates.
We now evaluate these expressions in the specific case $F=\left(\frac{u}{l}\right)^{d-1}$, $G=\left(\frac{u}{l}\right)^{d-3}$--see eq.(\ref{EEgeneric}).
This implies $F(u_0)=\left(\frac{u_0}{l}\right)^{d-1}$, hence $u_0^{d-1}=l^{d-1}\sqrt{c_y^2- c^2_t}$, which yields
\begin{eqnarray}
& &  T=2 c_t ~l^{d+1} \int_{u_0}^{\infty}\frac{du}{u^2}\frac{1}{\sqrt{u^{2(d-1)}-u_0^{2(d-1)}}},\label{Tads}\\
& &Y=2 c_y~ l^{d+1} \int_{u_0}^{\infty}\frac{du}{u^2}\frac{1}{\sqrt{u^{2(d-1)}-u_0^{2(d-1)}}}.\nonumber
\end{eqnarray}
For the EE we find,
\begin{eqnarray}
\frac{4 G_{10}}{\cal N}S_{EE}&=&    \frac{2}{l^{d-3}}\Bigg[\int_{u_0}^\infty du \frac{u^{2d-4}}{\sqrt{u^{2d-2} - u_0^{2d-2} }}  \nonumber\\
& &~~~~~~~~~- \int_0^\infty u^{d-3}~ du \Bigg].\label{SEEAdS}
\end{eqnarray}
At this point we observe that there are {\it two qualitatively different situations}. We refer to them as Type I and Type II embeddings. These are characterised by the value of
\begin{equation}
u_0= l\left( c_y^2-c_t^2\right)^{\frac{1}{2(d-1)}}.\label{u0gen}
\end{equation}
The  Type I embeddings have $(c_y^2-c_t^2)>0$. The turning point occurs for a real value of the coordinate $u=u_0$ in eq.(\ref{u0gen}). In contrast, for Type II embeddings (when $c_t^2>c_y^2$) we have
\begin{equation}
u_0= le^{\frac{i\pi}{2(d-1)}}\left[\sqrt{|c_y^2-c_t^2|}\right]^{\frac{1}{d-1}}= l \left( i \tilde{u}_0\right)^{\frac{1}{d-1}}.   
\end{equation}
The turning point is at a complex value of the $u$-coordinate, note that $\tilde{u}_0 =\sqrt{|c_y^2-c_t^2|}$ is real.
\\
In the case in which $c_t=0$ we are calculating the usual Ryu-Takayanagi EE, and this is always a Type I embedding. For the case  $c_y=0$ we are considering the time-like EE of \cite{Doi:2023zaf, Doi:2022iyj} and we have a Type II embedding.
Turning on both $c_y$ and $c_t$ we are describing a situation that interpolates between the usual Euclidean Ryu-Takayanagi and the time-like case. Based on the above arguments, one should therefore think of \eqref{regulatedEE} as in general the entanglement entropy of an ``interpolating slab'' ($S_{EE}=S_{\text{interpolating}}$) that interpolates between a pure spacelike slab ($c_t=0$) and a pure timelike slab ($c_y=0$) depending on the choice of the parameters $c_y$ and $c_t$. We refer the reader to Figure \ref{fig:light-cone}.

It is nice to make contact with the ideas of \cite{Heller:2024whi, Guo:2025pru}. To deal with the integrals in eqs.(\ref{Tads}),(\ref{SEEAdS}), we introduce a new variable $r=\frac{u_0}{u}$. Notice that the coordinate $r$ is real for Type I embeddings and complex for Type II ones (assuming that $u$ is real-valued). For Type II embeddings we must give a prescription to deal with the branch-cut in the square-root as done in \cite{Guo:2025pru}. Conversely, one may think that $u$ is complex whilst $r$ is real and we are in the situation described by \cite{Heller:2024whi}.
The expressions for the $T$-separation and $Y$-separation are, 
\begin{eqnarray}
& &  T=\frac{2 c_t~ ~l^{d+1}}{u^d_0} I_1,~~~Y= \frac{2 c_y~~ l^{d+1}}{u_0^d} I_1,~\text{where}\nonumber\\
& &  I_1=\int_0^1 dr \frac{r^{d-1}}{\sqrt{1-r^{2(d-1)}}}=\frac{\sqrt{\pi } \Gamma \left(\frac{d}{2 d-2}\right)}{\Gamma \left(\frac{1}{2 (d-1)}\right)}.\label{YT-ads}
\end{eqnarray}
For the EE in eq.(\ref{SEEAdS}) we find,
\begin{equation}
 \frac{2 G_{10} ~l^{d-3}}{{\cal N}~ u_0^{d-2}}S_{EE}= I_2-I_3.\label{SEEAdS2}   
\end{equation}
The quantities $I_2,I_3$ explicitly read
\begin{eqnarray}
& & I_2= \int_\epsilon^1 \frac{dr}{r^{d-1}\sqrt{1- r^{2d-2}}}\nonumber\\
& & = -\frac{1}{(d-2)r^{d-2}} \, _2F_1\left(\frac{1}{2},\frac{2-d}{2 d-2};\frac{d}{2 (d-1)};r^{2 d-2}\right) \Bigg|_\epsilon^1,\nonumber\\
& & I_3= \int_{\epsilon}^\infty \frac{dr}{r^{d-1}}=\frac{1}{(d-2)~\epsilon^{d-2}}.\nonumber
\end{eqnarray}
We introduced the small parameter $\epsilon\to 0$ to UV-regulate the quantities $I_2, I_3$. One can check that the divergence (for $\epsilon\to 0$) in $I_2$ is precisely equal to the divergence  in $I_3$. This is the logic of the UV-regulation  in eq.(\ref{regulatedEE}). The result is
\begin{eqnarray}
 I_2-I_3=\frac{1}{(2-d)} ~{}_2F_1\left(\frac{1}{2}, \frac{2-d}{2d-2};\frac{d}{2d-2};1 \right).  \label{SEEI2I3} 
\end{eqnarray}
It is useful to write the entanglement entropy in terms of the physical quantities in the field theory, namely the separations $Y$ and $T$. To do this we find the integration constants $(c_y,c_t)$ from eq.(\ref{YT-ads}) and put this together with eq.(\ref{u0gen}) to obtain,
\begin{eqnarray}
& & c_t= \frac{T}{2 l^{d+1} I_1} u_0^d,~~ c_y= \frac{Y}{2 l^{d+1} I_1} u_0^d,\nonumber\\
& & u_0=\frac{2 l^{d+1} I_1}{\sqrt{Y^2-T^2}}.\label{u0generico}   
\end{eqnarray}
Using eq.(\ref{SEEAdS2}) gives,
\begin{equation}
S_{EE}= \frac{{\cal N} (I_2-I_3) 2^{d-3} l^{(d-1)^2} I_1^{d-2}}{G_{10}} \times \frac{1}{\left( Y^2-T^2\right)^{\frac{(d-2)}{2}}}   .\label{SEEAdSfinal}
\end{equation}
 If the interval  $\Delta^2= Y^2- T^2$ is positive we are considering Type I embeddings (as $c_y>c_t$), we find a real result (see  Figure \ref{fig:light-cone}). For the case of negative  $\Delta^2$ (that corresponds to $c_y < c_t$), we are considering Type II embeddings and the result is,
 \begin{equation}
S_{EE}= \frac{{\cal N} (I_2-I_3) 2^{d-3} l^{(d-1)^2} I_1^{d-2}}{G_{10}} \times \frac{e^{-i\pi\frac{(d-2)}{2}}}{ |Y^2-T^2|^{\frac{(d-2)}{2}}}   .\label{SEEAdSfinal2}
\end{equation}
Two special cases are of interest. In the case $c_t=0$, we are in the pure Ryu-Takayanagi case
with $S_{EE}\propto \frac{1}{Y^{d-2}}$. On the other hand, for $c_y=0$, we are in the case of a Type II embedding and we find $S_{EE}\propto \frac{e^{-i\pi\frac{(d-2)}{2}}}{|T|^{d-2}} $, reproducing the result in \cite{Doi:2023zaf, Nunez:2025gxq}. Notice that the result of (\ref{SEEAdSfinal2}) is imaginary for odd $d$. The expressions in eqs.(\ref{SEEAdSfinal}) and (\ref{SEEAdSfinal2}) precisely match those in \cite{Nunez:2025gxq} and \cite{Doi:2023zaf} after setting $Y=0$. Let us analyse a couple of interesting cases.

\underline{\bf The case of AdS$_3$}
\\AdS$_3$ (that is $d=2$) is specially interesting. Using eqs.(\ref{Tads}),(\ref{SEEAdS}) we have,
\begin{eqnarray}
& & \frac{T}{2c_t l^3}=\frac{Y}{2c_y l^3}=\int_{u_0}^\infty ~ du \frac{1}{u^2\sqrt{u^2-u_0^2}} =\frac{1}{u_0^2}.\label{TYAdS3}\\
& &\frac{4G_N}{{\cal N} l} S_{EE}= \int_{u_0}^\infty \frac{du}{\sqrt{u^2-u_0^2}}-\int_0^\infty \frac{du}{u}.\label{SEEAdS3} 
\end{eqnarray}
The integrals in eq.(\ref{TYAdS3}) yield the result in eq.(\ref{YT-ads}). The integral in eq.(\ref{SEEAdS3}) is more delicate and the result in eq.(\ref{SEEAdS2}) cannot be used (one could expect a  logarithm of $u_0$). In fact, to analyse the integral in (\ref{SEEAdS3}), it is convenient to introduce a UV-cutoff $\Lambda (\to\infty)$ and an IR cutoff $\frac{1}{\Lambda} (\to 0)$. We find 
\begin{eqnarray}
& &  \int_{u_0}^\Lambda \frac{du}{\sqrt{u^2-u_0^2}}-\int_{\frac{1}{\Lambda}}^\Lambda \frac{du}{u}=\nonumber\\
& & \log\Bigg[{u+\sqrt{u^2-u_0^2}}\Bigg]\Bigg|_{u_0}^\Lambda-\log\left[{\Lambda^2}\right]=\nonumber\\
& &=\log\left[\frac{\sqrt{Y^2-T^2}}{\Lambda~ l^3}\right] + O(\frac{u_0^2}{\Lambda^2}).\label{EEAdS3fin}
\end{eqnarray}
This reproduces the field theory result (derived using the replica trick) in \cite{Doi:2023zaf},
\\
\underline{\bf The null-separation case}
\\
In this case, $Y^2-T^2=0$, we encounter a  divergent result. We now ask how this divergence manifests. We consider the result in eqs.(\ref{SEEAdSfinal}) and (\ref{SEEAdSfinal2}) and write $Y^2-T^2=\Delta Y^+ \Delta Y^-$. The null case takes $\Delta Y^-\to 0$. We  scale the coordinates with a  parameter $\eta$ according to \cite{Bousso:2014uxa},
\begin{eqnarray}
& &u\to\frac{u}{\eta}, ~du\to \frac{du}{\eta}, ~dy^+\to dy^+, ~dy^-\to \eta^2 dy^-, \nonumber\\
& & d\vec{x}_{d-2}\to\eta ~d\vec{x}_{d-2},
\end{eqnarray}
where we defined $y^{\pm}=y\pm t$. The background in eq.(\ref{genericmetric}) is invariant and we have that S$_{EE}$ scales as
\begin{equation}
 S_{EE, null}   = \frac{{\cal N} (I_2-I_3) 2^{d-3} l^{(d-1)^2} I_1^{d-2}}{G_{10} |\Delta Y^+\Delta Y^-|^{\frac{(d-2)}{2}}} \times \frac{1}{ \eta^{d-2}} .\label{nullcase1}
\end{equation}
The null limit is $\eta\to 0$. In this way we observe that the entanglement entropy diverges as
$S_{EE}\propto \frac{1}{\eta^{d-2}}$, see also \cite{Bousso:2014uxa}.

If we are to interpolate between the results in eqs.(\ref{SEEAdSfinal}) and (\ref{SEEAdSfinal2}) we should prescribe how to deal with the divergence in eq.(\ref{nullcase1}). One possibility is to set $\sqrt{c_y^2-c_t^2}=i \epsilon$ (for some small number $\epsilon$), this moves $u_0$ into the complex plane, avoiding the  divergence and connecting the results in eqs.(\ref{SEEAdSfinal})-(\ref{SEEAdSfinal2}).

The analytic continuation of the turning point ($u_0$) along  the imaginary axis is equivalent to consider a complex extremal surface in a complex manifold where the radial direction $u$ is complex \cite{Heller:2024whi}. This has the advantage of constructing the usual Ryu-Takayanagi (RT) surface for pure time-like entanglement (which is complex for the present case) and checking the strong sub-additivity of tEE in a holographic set-up. There is a turning point ($u_0$) along the complex radial axis and the extremal surface ends on the (real) time-like slab at the boundary. On the other hand, in the disconnected picture of two space-like geodesics ending on a time-like geodesic \cite{Doi:2023zaf, Doi:2022iyj}, there is no turning point (as one is considering the AdS with real $u$ axis), and the RT surface is the union of space-like and time-like surfaces. The junction conditions (the point where the space-like geodesic meets the time-like one) are important and need to be imposed in those points. This makes the holographic proof of strong sub-additivity nontrivial.
\\
Let us now analyse the same type of calculation in a simple confining model.

%
%
%
%
\section{Calculation in a  confining system}
We now consider the same observable calculated for the holographic dual to a field theory that is conformal and four dimensional at high energies, which spontaneously compactifies to a three-dimensional theory,  confining at long distances. The  background is written in \cite{Anabalon:2021tua}. Similar models were further analysed in \cite{Anabalon:2022aig, Anabalon:2024che, Chatzis:2024kdu, Chatzis:2024top, Chatzis:2025dnu, Castellani:2024ial, Giliberti:2024eii, Fatemiabhari:2024aua, Kumar:2024pcz, Nunez:2023xgl, Barbosa:2024smw, Nunez:2023nnl, Macpherson:2025pqi}. It would be interesting to perform the calculation in those models.

The background consists of a metric, a constant dilaton and a Ramond five form (not quoted here). The metric reads,
\begin{eqnarray}
& & ds_{10}^2= \frac{u^2}{l^2}\left[\lambda dt^2+dy^2+dx^2 + f(u) d\phi^2 \right]+\frac{l^2 du^2}{f(u) u^2}\nonumber\\
& &+ l^2 d\tilde{\Omega}_5^2,\label{ARmetric}\\
& & d\tilde{\Omega}_5^2= d\theta^2+ \sin^2\theta d\psi^2 + \sin^2\theta \sin^2\psi\left(d\varphi_1-A_1 \right)^2+\nonumber\\
& &~~~~~~~\sin^2\theta \cos^2\psi\left(d\varphi_2-A_1 \right)^2 +\cos^2\theta \left(d\varphi_3-A_1 \right)^2.\nonumber\\
& & A_1= Q\left(1-\frac{l^2Q^2}{u^2} \right)d\phi,~f(u)= 1-\left(\frac{Ql}{u}\right)^6.\nonumber
\end{eqnarray}
As above $\lambda=\pm 1$, with our focus being mostly on the Lorentzian case $\lambda=-1$. The $u$-coordinate ranges in $[u_\Lambda,\infty)$, the space ending smoothly  at $u_\Lambda=Ql$, for a particular period $L_\phi=\frac{1}{3Q}$ of the $\phi$-coordinate.
Consider the embedding of the eight manifold $\Sigma_8=[x,\phi,\tilde{\Omega}_5, u]$ with $t=t(u)$ and $y=y(u)$. The induced metric on $\Sigma_8$ is, 
\begin{eqnarray}
 & &   ds^2_8=\Big[ \frac{u^2}{l^2}(\lambda t'^2+y'^2)+\frac{l^2}{u^2 f}\Big]du^2 +\frac{u^2}{l^2}dx^2 + \frac{u^2}{l^2}f(u)d\phi^2 \nonumber\\
 & &+l^2 d\tilde{\Omega}_5^2.
\end{eqnarray}
The entanglement entropy reads,
\begin{eqnarray}
   & & S^{(\lambda)}_{EE}=\frac{\hat{\mathcal{N}}}{4G_{10}}\int_{u_0}^\infty \sqrt{G^2 (u)+F^2(u)(\lambda t'^2(u)+y'^2(u))},\label{EEconf}\nonumber\\
    & & G(u)=\frac{u}{l}, F(u)=\frac{u^3}{l^3}\sqrt{f(u)},\\
    & &\hat{\mathcal{N}}=l^5 L_x L_{\phi}\text{Vol}(\tilde{S}^5), L_\phi=\frac{1}{3Q}.\nonumber
\end{eqnarray}
Here, $\hat{\cal N}$ is related to the four dimensional UV-CFT central charge. In what follows we focus on the Lorentzian case ($\lambda=-1$). Using the generic expressions in eqs.(\ref{eq3})-(\ref{regulatedEE}) we write
\begin{eqnarray}
  & &  \frac{T}{2 ~l^5~  c_t}\!=\!\frac{Y}{2~l^5~ c_y}\!=\! 
\int_{u_0}^{\infty} \frac{u du}{\sqrt{u^6 -u_\Lambda^6}\sqrt{u^6 -u^6_0}}.\label{YT-confining}
\end{eqnarray}
The regularised entanglement entropy reads,
\begin{eqnarray}
  & &   \frac{2 ~l ~G_{10}}{\hat{\mathcal{N}}}S_{EE}=\int_{u_0}^{\infty} du\frac{u \sqrt{u^6 -u^6_\Lambda} }{\sqrt{u^6 -u^6_0}}- \int_{u_\Lambda}^{\infty}\! udu.\label{EEregconf}
\end{eqnarray}
We remind that  $u_{\Lambda}=Ql$ is the IR end of the space and  $u_0$ is the turning point defined as
\begin{align}
    u_0=l(Q^6+c^2_y -c^2_t)^{1/6}.\label{u0conf}
\end{align} 
It is instructive to compare the expressions above with those for CFTs in the previous section. In fact, for $d=4$ and $u_\Lambda=0$ (the parameter $u_\Lambda=Q l$ controls the presence of confinement), one finds that eq.(\ref{YT-confining}) reduces to eq.(\ref{Tads}), eq.(\ref{EEregconf}) boils-down to eq.(\ref{SEEAdS}) and eq.(\ref{u0conf}) to eq.(\ref{u0gen}). As above, we have two types of embedding depending on the sign of $(Q^6+c^2_y -c^2_t)$. Notice that in contrast with the conformal case, we can set $c_t>c_y$ and still have a real-valued turning point $u_0$ (a Type I embedding), by virtue of the parameter $Q>(c_t^2-c_y^2)^{1/6}$. This is an important observation of this work. Of course, for $c_t>\sqrt{Q^6+c_y^2}$, we have an imaginary value for $u_0=l ~e^{i\frac{\pi}{6}}|Q^6+c_y^2-c_t^2|^{\frac{1}{6}}$ and a Type II embedding for $\Sigma_8$.

Following similar steps as in the conformal case, we define,
\begin{eqnarray}
& &r=\frac{u_0}{u},~~~\gamma=\frac{u_\Lambda}{u_0},\label{definitionsconf} \\
& &J_1=\int_0^1 dr \frac{r^3}{\sqrt{(1-r^6)(1-\gamma^6 r^6)}}\nonumber\\
& &= \frac{\sqrt{\pi} ~\Gamma(\frac{5}{3})}{4 ~\Gamma(\frac{7}{6})} ~{}_2F_1[\frac{1}{2}, \frac{2}{3}, \frac{7}{6}, \gamma^6].\nonumber\\
& &J_2= \int_{\epsilon}^1\frac{dr}{r^3}\sqrt{\frac{1-\gamma^6 r^6}{1-r^6}}={\cal J}_2 +\frac{1}{2\epsilon^2}+O(\epsilon^4)\nonumber\\
& &{\cal J}_2= -\frac{\sqrt{\pi} \Gamma[\frac{2}{3}]}{2\Gamma[\frac{1}{6}]} ~{}_2F_1[-\frac{1}{2},-\frac{1}{3}, \frac{1}{6},\gamma^6],\nonumber\\
& &J_3= \int_{\epsilon}^{\frac{1}{\gamma}} \frac{dr}{r^3}= \frac{1}{2\epsilon^2}-\frac{\gamma^2}{2}.\nonumber
\end{eqnarray}

Using these definitions and the change of variables in eq.(\ref{definitionsconf}), we find
\begin{eqnarray}
& &  c_t=\frac{T~u_0^4}{2 J_1 ~l^5}, ~~~ c_y=  \frac{Y~u_0^4}{2 J_1 ~l^5}, \nonumber\\
& &\frac{2 ~l ~G_{10}}{\hat{\mathcal{N}} u_0^2}S_{EE}= J_2-J_3. \label{TY-confining}
\end{eqnarray}
Which should be compared with the analog expressions for CFT in four-dimensions, keeping in mind that $J_1$, $J_2$ and $J_3$ do depend on $u_0$.

It would be natural to find an expression for $S_{EE}$ in terms of the squared-interval $(Y^2-T^2)$ as we did in the conformal case in eqs.(\ref{u0generico})-(\ref{SEEAdSfinal2}). This is not simple in this case as the parameter $\gamma$ does depend on $u_0$ and so do the quantities $J_1,J_2,J_3$.  We can only express the result implicitly as,
\begin{eqnarray}
& &\Delta^2=(Y^2-T^2)= (1-\gamma^6)\frac{4~ l^4~ J_1^2}{u_0^2}, \nonumber\\
& &S_{EE}= \frac{\hat{\cal N}}{2~ l~ G_{N}}(J_2 -J_3)~u_0^2.    
\end{eqnarray}

\begin{figure}
    \centering
    \includegraphics[width=0.8\linewidth]{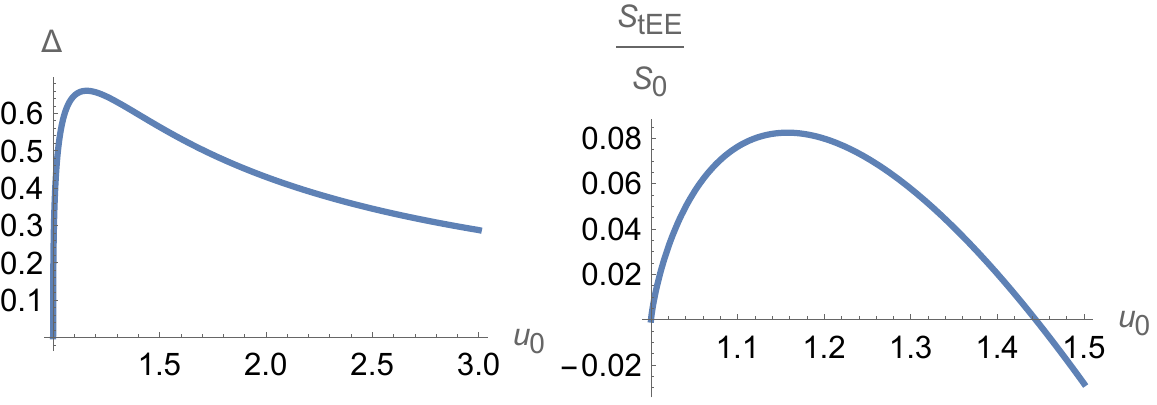}
    \caption{Plot of the interval $\Delta$ and the entanglement in terms of $u_0$,  with  $u_\Lambda=Q =1$.}
    \label{fig1}
\end{figure}

\begin{figure}
    \centering
    \includegraphics[width=0.5\linewidth]{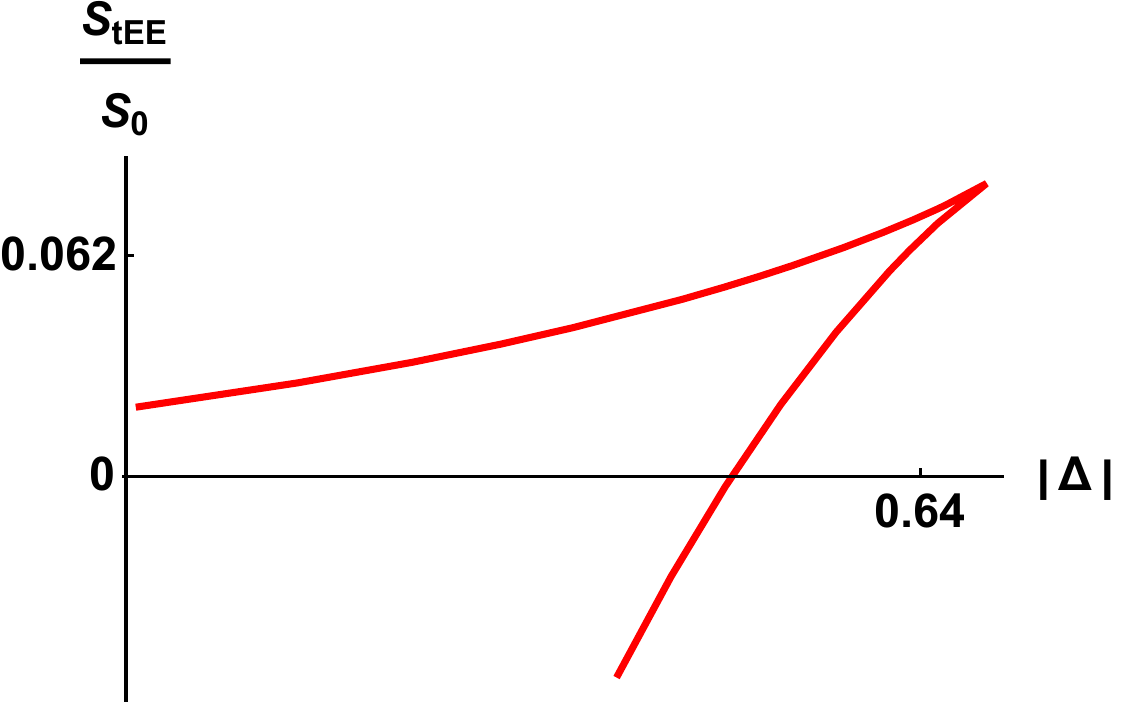}
    \caption{Parametric plot of $S_{tEE}$ vs. $\Delta$. We fix $u_\Lambda=Q=1$.}
    \label{fig3}
\end{figure}

Plots of $\Delta$ and $S_{EE}$ in terms of $u_0$ are shown in Figure \ref{fig1}. In  Figure \ref{fig3} we find a parametric plot of $S_{EE}$ in terms of $\Delta$. The non-monotonicity in the interval ($\Delta$) is indicative of phase transition in the EE (Figure \ref{fig1}). This is further ensured in Figure \ref{fig3}, which reveals a double value in EE for a given separation $\Delta$, which is typical of confining models and is indicative of a first-order phase transition \cite{Klebanov:2007ws, Kol:2014nqa, Jokela:2020wgs, Jokela:2025cyz, Barbosa:2024pyn}.

\section{Summary and conclusions}

In this work, we have investigated entanglement entropy  in strongly coupled quantum field theories with holographic duals, focusing on configurations that interpolate between space-like and time-like subsystem separations via Lorentz boosts. By studying boosted slab-like regions in both conformal and confining geometries, we identified two distinct classes of extremal surfaces: real (Type I) and complex (Type II), that together enable a unified treatment of entanglement across the full causal structure. These surfaces interpolate between the Ryu–Takayanagi  prescription and its time-like generalizations, providing a controlled setting to probe the interface between geometric and causal features of holographic EE. The interpolation between Type I and Type II embeddings is subtle and requires an analytic continuation to avoid the divergence in the case of null separation.

In fact, a salient outcome of our analysis is the emergence of complex extremal surfaces as a natural continuation of the entanglement functional into regimes that lie beyond the RT domain, particularly in the vicinity of the light cone. In conformal backgrounds, the EE exhibits a subtle interpolation across causal boundaries. In confining geometries, we observe first-order phase transitions. Another important observation made in the confining case is that we can have Type I surfaces, even for time-like separations in the dual QFT ($c_t^2>c_y^2$).
The appearance of these transitions signals nontrivial reorganisation of the entanglement structure, which we interpret as a geometric manifestation of underlying confinement effects in the dual field theory.

Importantly, our results underscore the necessity of including complex-valued extremal surfaces when extending the holographic EE framework to time-like or null-separated regions. Although their physical interpretation remains subtle, particularly in the absence of a direct replica-trick derivation, these saddles play a central role in ensuring analytic continuity of the EE functional. Their inclusion suggests that the causal structure of the boundary theory is encoded not only in the real geometry of the bulk but also in a more intricate analytic landscape of complex saddles that mediate entanglement beyond traditional kinematic regimes.

Several directions merit further exploration. One natural extension involves examining boosted entanglement in more general bulk backgrounds, including finite temperature black branes, geometries associated with charged and rotating branes, including higher-derivative corrections, or  backgrounds dual to phenomenologically interesting QFTs  \cite{Casero:2006pt, Casero:2007jj, Hoyos-Badajoz:2008znk}. This might link with interesting Physics. Additionally, a deeper field-theoretic understanding of the role of complex saddles (perhaps via an analytically continued replica method or non-unitary deformations) could offer insight into the precise encoding of causality in entanglement data. More broadly, our work points toward a unified framework in which entanglement serves not only as a probe of spatial geometry but also as a dynamical bridge across distinct causal domains in holography.
These results offer new insights into the nature of time-like entanglement and provide a controlled setting to study analytic continuations of extremal surfaces in holography.

\bibliographystyle{unsrt}
\bibliography{refs}

\begin{thebibliography}{10}

\bibitem{Maldacena:1997re}
Juan~Martin Maldacena.
\newblock {The Large $N$ limit of superconformal field theories and
  supergravity}.
\newblock {\em Adv. Theor. Math. Phys.}, 2:231--252, 1998.

\bibitem{Ryu:2006bv}
Shinsei Ryu and Tadashi Takayanagi.
\newblock {Holographic derivation of entanglement entropy from AdS/CFT}.
\newblock {\em Phys. Rev. Lett.}, 96:181602, 2006.

\bibitem{Ryu:2006ef}
Shinsei Ryu and Tadashi Takayanagi.
\newblock {Aspects of Holographic Entanglement Entropy}.
\newblock {\em JHEP}, 08:045, 2006.

\bibitem{Hubeny:2007xt}
Veronika~E. Hubeny, Mukund Rangamani, and Tadashi Takayanagi.
\newblock {A Covariant holographic entanglement entropy proposal}.
\newblock {\em JHEP}, 07:062, 2007.

\bibitem{Swingle:2009bg}
Brian Swingle.
\newblock {Entanglement Renormalization and Holography}.
\newblock {\em Phys. Rev. D}, 86:065007, 2012.

\bibitem{VanRaamsdonk:2010pw}
Mark Van~Raamsdonk.
\newblock {Building up spacetime with quantum entanglement}.
\newblock {\em Gen. Rel. Grav.}, 42:2323--2329, 2010.

\bibitem{Nakata:2020luh}
Yoshifumi Nakata, Tadashi Takayanagi, Yusuke Taki, Kotaro Tamaoka, and Zixia
  Wei.
\newblock {New holographic generalization of entanglement entropy}.
\newblock {\em Phys. Rev. D}, 103(2):026005, 2021.

\bibitem{Mollabashi:2021xsd}
Ali Mollabashi, Noburo Shiba, Tadashi Takayanagi, Kotaro Tamaoka, and Zixia
  Wei.
\newblock {Aspects of pseudoentropy in field theories}.
\newblock {\em Phys. Rev. Res.}, 3(3):033254, 2021.

\bibitem{Muller-Hermes:2012irk}
Alexander M{\"u}ller-Hermes, J.~Ignacio Cirac, and Mari~Carmen Ba{\~n}uls.
\newblock {Tensor network techniques for the computation of dynamical
  observables in one-dimensional quantum spin systems}.
\newblock {\em New J. Phys.}, 14(7):075003, 2012.

\bibitem{Hastings:2014qqa}
M.~B. Hastings and R.~Mahajan.
\newblock {Connecting Entanglement in Time and Space: Improving the Folding
  Algorithm}.
\newblock {\em Phys. Rev. A}, 91(3):032306, 2015.

\bibitem{Lerose:2020fhd}
Alessio Lerose, Michael Sonner, and Dmitry~A. Abanin.
\newblock {Influence Matrix Approach to Many-Body Floquet Dynamics}.
\newblock {\em Phys. Rev. X}, 11(2):021040, 2021.

\bibitem{Bou-Comas:2024pxf}
Aleix Bou-Comas, Carlos~Ramos Marim{\'o}n, Jan~T. Schneider, Stefano Carignano,
  and Luca Tagliacozzo.
\newblock {Measuring temporal entanglement in experiments as a hallmark for
  integrability}.
\newblock 9 2024.

\bibitem{Doi:2022iyj}
Kazuki Doi, Jonathan Harper, Ali Mollabashi, Tadashi Takayanagi, and Yusuke
  Taki.
\newblock {Pseudoentropy in dS/CFT and Timelike Entanglement Entropy}.
\newblock {\em Phys. Rev. Lett.}, 130(3):031601, 2023.

\bibitem{Doi:2023zaf}
Kazuki Doi, Jonathan Harper, Ali Mollabashi, Tadashi Takayanagi, and Yusuke
  Taki.
\newblock {Timelike entanglement entropy}.
\newblock {\em JHEP}, 05:052, 2023.

\bibitem{Heller:2024whi}
Michal~P. Heller, Fabio Ori, and Alexandre Serantes.
\newblock {Geometric Interpretation of Timelike Entanglement Entropy}.
\newblock {\em Phys. Rev. Lett.}, 134(13):131601, 2025.

\bibitem{Xu:2024yvf}
Jin Xu and Wu-zhong Guo.
\newblock {Imaginary part of timelike entanglement entropy}.
\newblock {\em JHEP}, 02:094, 2025.

\bibitem{Guo:2025pru}
Wu-zhong Guo and Jin Xu.
\newblock {A duality of Ryu-Takayanagi surfaces inside and outside the
  horizon}.
\newblock 2 2025.

\bibitem{Das:2023yyl}
Avijit Das, Shivrat Sachdeva, and Debajyoti Sarkar.
\newblock {Bulk reconstruction using timelike entanglement in (A)dS}.
\newblock {\em Phys. Rev. D}, 109(6):066007, 2024.

\bibitem{Grieninger:2023knz}
Sebastian Grieninger, Kazuki Ikeda, and Dmitri~E. Kharzeev.
\newblock {Temporal entanglement entropy as a probe of renormalization group
  flow}.
\newblock {\em JHEP}, 05:030, 2024.

\bibitem{Chu:2019uoh}
Chong-Sun Chu and Dimitrios Giataganas.
\newblock {$c$-Theorem for Anisotropic RG Flows from Holographic Entanglement
  Entropy}.
\newblock {\em Phys. Rev. D}, 101(4):046007, 2020.

\bibitem{Afrasiar:2024lsi}
Mir Afrasiar, Jaydeep~Kumar Basak, and Dimitrios Giataganas.
\newblock {Timelike entanglement entropy and phase transitions in non-conformal
  theories}.
\newblock {\em JHEP}, 07:243, 2024.

\bibitem{Afrasiar:2024ldn}
Mir Afrasiar, Jaydeep~Kumar Basak, and Dimitrios Giataganas.
\newblock {Holographic Timelike Entanglement Entropy in Non-relativistic
  Theories}.
\newblock 11 2024.

\bibitem{Milekhin:2025ycm}
Alexey Milekhin, Zofia Adamska, and John Preskill.
\newblock {Observable and computable entanglement in time}.
\newblock 2 2025.

\bibitem{Li:2022tsv}
Ze~Li, Zi-Qing Xiao, and Run-Qiu Yang.
\newblock {On holographic time-like entanglement entropy}.
\newblock {\em JHEP}, 04:004, 2023.

\bibitem{Roychowdhury:2025ukl}
Dibakar Roychowdhury.
\newblock {Holographic timelike entanglement and $c$ theorem for supersymmetric
  QFTs in ($ 0+1 $)d}.
\newblock 2 2025.

\bibitem{Nunez:2025gxq}
Carlos Nunez and Dibakar Roychowdhury.
\newblock {Time-like Entanglement Entropy: a top-down approach}.
\newblock 5 2025.

\bibitem{Giataganas:2025div}
Dimitrios Giataganas.
\newblock {Holographic Timelike c-function}.
\newblock 5 2025.

\bibitem{Bousso:2014uxa}
Raphael Bousso, Horacio Casini, Zachary Fisher, and Juan Maldacena.
\newblock {Entropy on a null surface for interacting quantum field theories and
  the Bousso bound}.
\newblock {\em Phys. Rev. D}, 91(8):084030, 2015.

\bibitem{Anabalon:2021tua}
Andres Anabalon and Simon~F. Ross.
\newblock {Supersymmetric solitons and a degeneracy of solutions in AdS/CFT}.
\newblock {\em JHEP}, 07:015, 2021.

\bibitem{Klebanov:2007ws}
Igor~R. Klebanov, David Kutasov, and Arvind Murugan.
\newblock {Entanglement as a probe of confinement}.
\newblock {\em Nucl. Phys. B}, 796:274--293, 2008.

\bibitem{Kol:2014nqa}
Uri Kol, Carlos N\'u\~nez, Daniel Schofield, Jacob Sonnenschein, and Michael
  Warschawski.
\newblock {Confinement, Phase Transitions and non-Locality in the Entanglement
  Entropy}.
\newblock {\em JHEP}, 06:005, 2014.

\bibitem{Heller:2025kvp}
Michal~P. Heller, Fabio Ori, and Alexandre Serantes.
\newblock {Temporal Entanglement from Holographic Entanglement Entropy}.
\newblock 7 2025.

\bibitem{Anabalon:2022aig}
Andr\'es Anabal\'on, Antonio Gallerati, Simon Ross, and Mario Trigiante.
\newblock {Supersymmetric solitons in gauged $ \mathcal{N} $ = 8 supergravity}.
\newblock {\em JHEP}, 02:055, 2023.

\bibitem{Anabalon:2024che}
Andr\'es Anabal\'on, Horatiu Nastase, and Marcelo Oyarzo.
\newblock {Supersymmetric AdS solitons and the interconnection of different
  vacua of $ \mathcal{N} $ = 4 Super Yang-Mills}.
\newblock {\em JHEP}, 05:217, 2024.

\bibitem{Chatzis:2024kdu}
Dimitrios Chatzis, Ali Fatemiabhari, Carlos Nunez, and Peter Weck.
\newblock {SCFT deformations via uplifted solitons}.
\newblock {\em Nucl. Phys. B}, 1006:116659, 2024.

\bibitem{Chatzis:2024top}
Dimitrios Chatzis, Ali Fatemiabhari, Carlos Nunez, and Peter Weck.
\newblock {Conformal to confining SQFTs from holography}.
\newblock {\em JHEP}, 08:041, 2024.

\bibitem{Chatzis:2025dnu}
Dimitrios Chatzis, Madison Hammond, Georgios Itsios, Carlos Nunez, and
  Dimitrios Zoakos.
\newblock {Universal Observables, SUSY RG-Flows and Holography}.
\newblock 6 2025.

\bibitem{Castellani:2024ial}
Federico Castellani and Carlos Nunez.
\newblock {Holography for confined and deformed theories: TsT-generated
  solutions in type IIB supergravity}.
\newblock {\em JHEP}, 12:155, 2024.

\bibitem{Giliberti:2024eii}
Mauro Giliberti, Ali Fatemiabhari, and Carlos Nunez.
\newblock {Confinement and screening via holographic Wilson loops}.
\newblock {\em JHEP}, 11:068, 2024.

\bibitem{Fatemiabhari:2024aua}
Ali Fatemiabhari and Carlos Nunez.
\newblock {From conformal to confining field theories using holography}.
\newblock {\em JHEP}, 03:160, 2024.

\bibitem{Kumar:2024pcz}
S.~Prem Kumar and Ricardo Stuardo.
\newblock {Twisted circle compactification of $ \mathcal{N} $ = 4 SYM and its
  holographic dual}.
\newblock {\em JHEP}, 08:089, 2024.

\bibitem{Nunez:2023xgl}
Carlos Nunez, Marcelo Oyarzo, and Ricardo Stuardo.
\newblock {Confinement and D5-branes}.
\newblock {\em JHEP}, 03:080, 2024.

\bibitem{Barbosa:2024smw}
Marcelo Barbosa, Horatiu Nastase, Carlos Nunez, and Ricardo Stuardo.
\newblock {Penrose limits of I-branes, twist-compactified D5-branes, and spin
  chains}.
\newblock {\em Phys. Rev. D}, 110(4):046015, 2024.

\bibitem{Nunez:2023nnl}
Carlos Nunez, Marcelo Oyarzo, and Ricardo Stuardo.
\newblock {Confinement in (1 + 1) dimensions: a holographic perspective from
  I-branes}.
\newblock {\em JHEP}, 09:201, 2023.

\bibitem{Macpherson:2025pqi}
Niall Macpherson, Paul Merrikin, Carlos Nunez, and Ricardo Stuardo.
\newblock {Twisted-Circle Compactifications of SQCD-like Theories and
  Holography}.
\newblock 6 2025.

\bibitem{Jokela:2020wgs}
Niko Jokela and Javier~G. Subils.
\newblock {Is entanglement a probe of confinement?}
\newblock {\em JHEP}, 02:147, 2021.

\bibitem{Jokela:2025cyz}
Niko Jokela, Jani Kastikainen, Carlos Nunez, Jos\'e~Manuel Pen\'\i{}n, Helime
  Ruotsalainen, and Javier~G. Subils.
\newblock {On entanglement c-functions in confining gauge field theories}.
\newblock 5 2025.

\bibitem{Barbosa:2024pyn}
Sergio Barbosa, Sylvain Fichet, Eugenio Megias, and Mariano Quiros.
\newblock {Entanglement and Thermal Transitions from Singularities}.
\newblock 6 2024.

\bibitem{Casero:2006pt}
Roberto Casero, Carlos N\'u\~nez, and Angel Paredes.
\newblock {Towards the string dual of N=1 SQCD-like theories}.
\newblock {\em Phys. Rev. D}, 73:086005, 2006.

\bibitem{Casero:2007jj}
Roberto Casero, Carlos Nunez, and Angel Paredes.
\newblock {Elaborations on the String Dual to N=1 SQCD}.
\newblock {\em Phys. Rev. D}, 77:046003, 2008.

\bibitem{Hoyos-Badajoz:2008znk}
Carlos Hoyos-Badajoz, Carlos Nunez, and Ioannis Papadimitriou.
\newblock {Comments on the String dual to N=1 SQCD}.
\newblock {\em Phys. Rev. D}, 78:086005, 2008.

\end{thebibliography}

\noindent \textit{Acknowledgments:} We thank Wu-zhong Guo, Prem Kumar,  Michal Heller, Fabio Ori, Simon Ross, Alexandre Serantes, Tadashi Takayanagi
for useful and interesting comments. DR would like to acknowledge The Royal Society, UK for financial assistance. DR also acknowledges the Mathematical Research Impact Centric Support (MATRICS) grant (MTR/2023/000005) received from ANRF, India. C. N. is supported by STFC’s grants ST/Y509644-1, ST/X000648/1 and ST/T000813/1.

\end{document}